\documentclass[aps, reprint, superscriptaddress, prl, preprintnumbers, amsmath, amssymb, letterpaper, floatfix, showpacs]{revtex4-1}

\usepackage{amsmath, amssymb}
\usepackage{graphicx}
\usepackage{times}
\usepackage{dcolumn}
\newcommand{\msr}{$\mu$SR}
\newcommand{\YBCO}{YBa$_{2}$Cu$_{3}$O$_{y}$}
\newcommand{\YBCOMS}{YBa$_{2}$Cu$_{3}$O$_{y}$, $y = 6.77$ and 6.83}
\newcommand{\YBCOS}{YBa$_{2}$Cu$_{3}$O$_{6.77}$}
\newcommand{\YBCOE}{YBa$_{2}$Cu$_{3}$O$_{6.83}$}%
\newcommand{\BSCCO}{Bi$_{2+x}$Sr$_{2-x}$CaCu$_{2}$O$_{8+\delta}$}

\begin{document}

\preprint{}

\title{\boldmath Muon Spin Relaxation and fluctuating magnetism in the pseudogap phase of \YBCO}
\author{Z. H. Zhu}
\altaffiliation[ ]{These authors contributed equally to this work.}
\author{J. Zhang}
\altaffiliation[ ]{These authors contributed equally to this work.}
\author{Z. F. Ding}
\author{C. Tan}
\author{C. S. Chen}
\author{Q. Wu}
\author{Y. X. Yang}
\affiliation{State Key Laboratory of Surface Physics, Department of Physics, Fudan University, Shanghai 200433, People's Republic of China}
\author{O. O. Bernal}
\affiliation{Department of Physics and Astronomy, California State University, Los Angeles, CA 90032, USA}
\author{P.-C. Ho}
\affiliation{Department of Physics, California State University, Fresno, CA 93740, USA}
\author{G. D. Morris}
\affiliation{TRIUMF, Vancouver, British Columbia V6T 2A3, Canada}
\author{A. Koda}
\affiliation{Muon Science Laboratory and Condensed Matter Research Center, Institute of Materials Structure Science, High Energy Accelerator Research Organization (KEK), Tsukuba, Ibaraki 305-0801, Japan}
\affiliation{Department of Materials Structure Science, The Graduate University for Advanced Studies (Sokendai), Tsukuba, Ibaraki 305-0801, Japan}
\author{A. D. Hillier}
\author{S. P. Cottrell}
\author{P. J. Baker}
\author{P. K\@. Biswas}
\affiliation{ISIS Facility, STFC Rutherford Appleton Laboratory, Harwell Science and Innovation Campus, Chilton, Didcot, Oxon, UK}
\author{J. Qian}
\affiliation{Key Laboratory of Artificial Structures and Quantum Control (Ministry of Education), School of Physics and Astronomy, Shanghai Jiao Tong University, Shanghai 200240, People's Republic of China}
\author{X. Yao}
\affiliation{Key Laboratory of Artificial Structures and Quantum Control (Ministry of Education), School of Physics and Astronomy, Shanghai Jiao Tong University, Shanghai 200240, People's Republic of China}
\affiliation{Collaborative Innovation Center of Advanced Microstructures, Nanjing 210093, People's Republic of China}
\author{D. E. MacLaughlin}
\affiliation{Department of Physics and Astronomy, University of California, Riverside, Riverside, CA 92521, USA}
\author{L. Shu}
\altaffiliation[Corresponding Author: ]{leishu@fudan.edu.cn}
\affiliation{State Key Laboratory of Surface Physics, Department of Physics, Fudan University, Shanghai 200433, People's Republic of China}
\affiliation{Collaborative Innovation Center of Advanced Microstructures, Nanjing 210093, People's Republic of China}
\affiliation{Shanghai Research Center for Quantum Sciences, Shanghai 201315, People's Republic of China}

\date{\today}

\begin{abstract}
We report results of a muon spin relaxation study of slow magnetic fluctuations in the pseudogap phase of underdoped single-crystalline \YBCOMS. The dependence of the dynamic muon spin relaxation rate on applied magnetic field yields the rms magnitude~$B\mathrm{_{loc}^{rms}}$ and correlation time~$\tau_c$ of fluctuating local fields at muon sites. The observed relaxation rates do not decrease with decreasing temperature~$T$ below the pseudogap onset at $T^\ast$, as would be expected for a conventional magnetic transition; both $B\mathrm{_{loc}^{rms}}$ and $\tau_c$ are roughly constant in the pseudogap phase down to the superconducting transition. Corresponding NMR relaxation rates are estimated to be too small to be observable. Our results put strong constraints on theories of the anomalous pseudogap magnetism in \YBCO.
\end{abstract}

\maketitle

The pseudogap phase in hole-doped cuprates is one of the most studied quantum states in high-temperature superconductors~\cite{Emery1995, Timusk_1999, Lee1999, Kivelson03, Norman03, Basov2005, Review2005, CMV2010, NormanScience, Keimer2015}. It is characterized by the loss of low-lying electronic excitations, and emerges below a characteristic temperature $T^\ast$ that depends strongly on the hole concentration on the CuO$_{2}$ plane. Anomalous transport~\cite{LT2019}, thermodynamic~\cite{Loram1994, LT2019}, and electrodynamic~\cite{Basov2005} properties are observed below $T^\ast$. Extensive work~\cite{Timusk_1999} has shown that the pseudogap state is quite different from a normal metal, and it is widely believed that it holds the key to a general model for high-temperature superconductivity. Two categories of theories, involving either a crossover~\cite{Emery1995, Lee1999, Keimer2015} or a true thermodynamic phase transition with a quantum critical point~\cite{Kivelson03, CMV2010}, attempt to explain the origin of the pseudogap. Both are consistent with several experimental phenomena~\cite{Keimer2015}.

A variety of symmetry-sensitive techniques have discovered broken inversion and time-reversal symmetries (TRS) in a number of cuprate superconductors below $T^\ast$~\cite{Fauque2006, Li2008, Xia2008, ManginThro2015, Lubashevsky2014, Zhao2016, Sato2017, Zhang2018, Pal2018}. Among the various models, theories considering intra-unit-cell (IUC) magnetic order~\cite{CMV2002, Moskvin2012, Fechner2016} have been proposed in which both of these symmetries are broken. Objections to IUC magnetism were raised, however, as probes of magnetic moments and local fields expected from TRS breaking yielded a wide variety of results. Polarized neutron diffraction experiments yielded evidence for TRS-breaking moments below $T^\ast$ of the order of 0.1$\mu_\mathrm{B}$~\cite{Fauque2006, Li2008, ManginThro2015}, but also for the absence of such moments~\cite{Croft2017}, although the latter has been questioned~\cite{Bourges2018, Croft2018}. Claims of both the presence~\cite{sonier2001science, Zhang2018, Pal2018} and absence~\cite{sonier2002diffusion, MacDougall2008, Sonier2009, Sonier2012, Gheidi2020} of static and/or dynamic fields in the pseudogap phase have been made based on muon spin relaxation (\msr) experiments~\cite{Brewer2003, Yaouanc2011} carried out using different configurations. NMR studies~\cite{Strassle2008, Strassle2011, Mounce2013} have not observed such fields. 

Our recent \msr\ measurements of dynamic muon relaxation rates~$\lambda$ in \YBCO~\cite{Zhang2018} revealed slowly-fluctuating magnetic fields $B_\mathrm{loc}(t)$, with heuristic estimates of the root-mean-square (rms) magnitude~$B\mathrm{_{loc}^{rms}} \equiv \langle B_\mathrm{loc}^2(t)\rangle^{1/2} \approx 1$~mT and correlation time~$\tau_{c} \approx 10$~ns. Such fluctuations are consistent with polarized neutron diffraction results, for which the experimental time scale (${\sim}10^{-12}$~s) is much shorter than $\tau_c$. They also explain the absence of static fields in NMR and \msr\ experiments, where the time scale ($\gtrsim 10^{-5}$~s) is considerably longer. It has been suggested~\cite{CMV2014} that the fluctuations arise from quantum size effects in domains of IUC order.

We also observed maxima in $\lambda(T)$ at temperatures~$T_\mathrm{mag} \approx T^\ast$ in \YBCO, $y = \text{6.72}$, 6.77, and 6.95, followed by increases of $\lambda$ with decreasing temperature in the pseudogap phase~\cite{Zhang2018}. It was determined that the maxima were not due to activated muon hopping, charge inhomogeneity, nuclear-dipole fields, or other phenomena, but were closely related with the formation of IUC magnetic order. A recent \msr\ study of \BSCCO~\cite{Pal2018} found quasistatic magnetic fluctuations in the pseudogap phase that might have the same origin as those in \YBCO.

Here we report improved measurements of $B\mathrm{_{loc}^{rms}}$ and $\tau_c$ in the pseudogap phase of \YBCOMS. These oxygen concentrations were chosen so that $T^\ast$ is above the onsets of charge density wave phases~\cite{Wu2015, Keimer2015} but well below the the muon hopping temperature regime~$T \gtrsim 200$~K~\cite{Zhang2018}. We determine $B\mathrm{_{loc}^{rms}}$ and $\tau_c$ separately by extracting them from the field dependence of $\lambda$ measured at a number of temperatures. We find that they are each roughly constant from just above $T^\ast$ down to the superconducting transition temperature~$T_c$. This is in contrast to ordinary magnetic transitions, where both quantities decrease with decreasing temperature in the ordered state~\footnote{See, for example, Ref.~\cite{Yaouanc2011}, Chap.~10}, and raises the question of the origin of the spin dynamics at low temperatures.

Parent compounds for the single crystalline samples of \YBCOMS, were synthesized by a polythermal top-seeded solution-growth method using a 3BaO-5CuO solvent flux~\cite{Xiang2016}. This method can yield crystals with high crystallinity~\cite{growth1996}. The bulk single crystal was then cut into small pieces with $ab$ plane lateral dimensions of 2~mm$\times$2~mm and $c$-axis thicknesses of 0.5 mm. Oxygen concentrations~$y$ = 6.77 and 6.83 were achieved by post-annealing the parent compound in flowing ultra-high-purity oxygen at different temperatures~\cite{Schleger1991, Gao2006, Zhang2018}. For comparison, values of $T_c$, $T^\ast$, and $T_\mathrm{mag}$ are given in Table~\ref{tab:temps}.
\begin{table} [ht]
\caption{\label{tab:temps} Superconducting transition temperatures~$T_c$, pseudogap onset temperatures~$T^\ast$, and peak temperatures~$T_\mathrm{mag}$ in $\mu^+$ ZF relaxation rates for \YBCOMS. From Ref.~\cite{Zhang2018} except as noted.}
\begin{ruledtabular}
\begin{tabular}{cccl}
$y$ & $T_c$ (K) (onset) & $T^\ast$ (K) (approx.) & $T_\mathrm{mag}$ (K) \\
\colrule
6.77 & 80 & 155--185 & 160(10) \\ 
6.83 & 88 & 130--160 & 142(10)~\footnote{Unpublished data.} \\ 
\end{tabular}
\end{ruledtabular}
\end{table}

\msr\ experiments were carried out using the LAMPF spectrometer at TRIUMF, Vancouver, Canada; the EMU spectrometer at the ISIS Facility, Rutherford Appleton Laboratory, Chilton, United Kingdom; and the ARTEMIS spectrometer at J-PARC, Tokai, Japan. At all facilities, 100\% positively-charged spin-polarized muons ($\mu^+$) were implanted into the samples with the initial ensemble $\mu^+$ spin polarization $\mathbf{P}_{\mu}(0)$ normal to the $ab$ plane.

Previous \msr\ experiments on \YBCO~\cite{Zhang2018} revealed that fluctuations of local fields at $\mu^+$ sites are motionally narrowed: $\gamma_\mu B\mathrm{_{loc}^{rms}}\tau_{c} \ll 1$, where $\gamma_{\mu}$ = $8.5156 \times 10^{8}~\mathrm{s}^{-1}~\mathrm{T}^{-1}$ is the muon gyromagnetic ratio. In an externally applied longitudinal field (LF)~$\mathbf{H}_L \parallel \mathbf{P}_{\mu}(0)$, the corresponding motion\-ally-narrowed relaxation rate~$\lambda_L$ follows the so-called Redfield relation~\cite{Hayano1979, Slichter1996}
\begin{equation}
\label{eq:Redfield}
\lambda_L(H_L) = \frac{2(\gamma_{\mu} B\mathrm{_{loc}^{rms}})^{2}\,\tau_{c}}{1+(\gamma_{\mu}\mu_0H_L\tau_{c})^{2}},
\end{equation}
if the fluctuations are Markovian and characterized by a single correlation time~$\tau_{c}$. Equation~(\ref{eq:Redfield}) represents the effect of sweeping the $\mu^+$ Zeeman frequency~$\gamma_\mu\mu_0H_L$ through the fluctuation noise spectrum, and assumes no field dependence of the spin dynamics. In Eq.~(\ref{eq:Redfield}) a crossover occurs for $\gamma_\mu\mu_0H_L \approx 1/\tau_c$, and the area~$\int_0^\infty \lambda_L(H_L)\,dH_L =\pi\gamma_\mu B_\mathrm{loc}^2$ is independent of $\tau_c$. For more general fluctuation spectra, $\lambda_L(H_L)$ decreases with increasing $H_L$ as the $\mu^+$ Zeeman frequency passes through a high-frequency cutoff. The Redfield relation has been widely applied in \msr\ to characterize dynamic fluctuating magnetic fields in strongly correlated electron systems~\cite{Aoki2003,Li2016,Tan2019}.

The observed $\mu^+$ spin relaxation in \YBCO\ is very slow~\cite{Zhang2018}, and care must be taken to confirm the absence of spurious spectrometer-dependent signals. $\mu^+$ relaxation is even slower in pure silver~\cite{Bueno2011, Zhang2018}, so that control experiments on Ag samples serve as a check for such signals. In the present study, LF-$\mu$SR data were taken on the LAMPF spectrometer at TRIUMF and the ARTEMIS spectrometer at J-PARC, using a pure silver sample with lateral dimension and thickness similar to those of the \YBCO\ samples. 

\begin{figure} [ht]
 \begin{center}
 \includegraphics[clip=,width=0.5\textwidth]{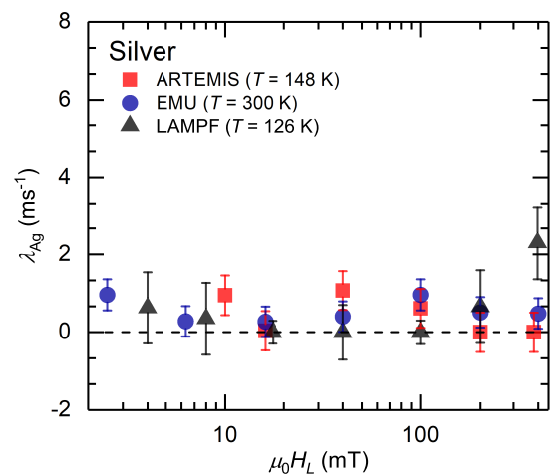}
 \caption{Longitudinal muon spin relaxation rates in pure silver samples with dimensions similar to those of the \YBCO\ samples. Data taken using three different spectrometers at different temperatures. Squares: J-PARC/ARTEMIS\@. Circles: ISIS/EMU (from Ref.~\cite{Zhang2018}). Triangles: TRIUMF/LAMPF.}
 \label{fig:Ag}
 \end{center}
\end{figure}

Figure~\ref{fig:Ag} 
shows the field dependence of $\mu^+$ dynamic relaxation rates~$\lambda_\mathrm{Ag}$ measured on the spectrometers used in this work. Data from the EMU spectrometer, taken from the supplementary information to Ref.~\cite{Zhang2018}, are also shown. The values~$\lambda_\mathrm{Ag} \approx 1~\text{ms}^{-1}$ are in good agreement with previous results~\cite{Bueno2011, Zhang2018}, and serve as a field-independent upper bound on any such signal up to 400~mT\@. 

We then measured $\lambda_L(H_L)$ in \YBCOMS, at various temperatures above the superconducting transitions at $T_c(y)$. All data were taken using the LAMPF spectrometer except for $y = 6.83$, $T = 170$~K, which were taken using the ARTEMIS spectrometer. 

For the LAMPF data two small spectrometer corrections were necessary because of the very slow relaxation rates~\footnote{(1)~In a muon beam for $\mu$SR the $\mu^+$ spin polarization is initially (anti)\-parallel to the beam direction, but must precess at least slightly as the beam passes through a separator (crossed electric and magnetic fields) that removes positron contamination from the beam. Thus there is always a small angle between the stopped $\mu^+$ initial spin direction and the applied field (which is parallel to the beam). The resulting $\mu^+$ precession at frequency~$\gamma_\mu\mu_0H_L$ would not contribute to the LF signal for perfect axial symmetry, but a small oscillating component is often observed. \par (2)~In the LAMPF cryostat the sample is suspended in a large ring, so that muons that miss the sample ``fly past'' and are vetoed by downstream counters. However, data taken with the sample removed revealed that a few percent of the muon flux stops in the sample holder and cryostat and relaxes with a significant rate. \par Corrections for both of these effects were incorporated into the fit functions.}. In the ARTEMIS cryostat a sizable contribution to the signal was observed from muons that miss the sample and stop in the silver sample holder. Its magnitude was determined using zero field (ZF) data, where the sample and Ag contributions could be separately determined because of their different relaxation rates. The correction for LF data included Ag rates from Fig.~\ref{fig:Ag}.

\begin{figure} [ht]
 \begin{center}
 \includegraphics[clip=,width=0.5\textwidth]{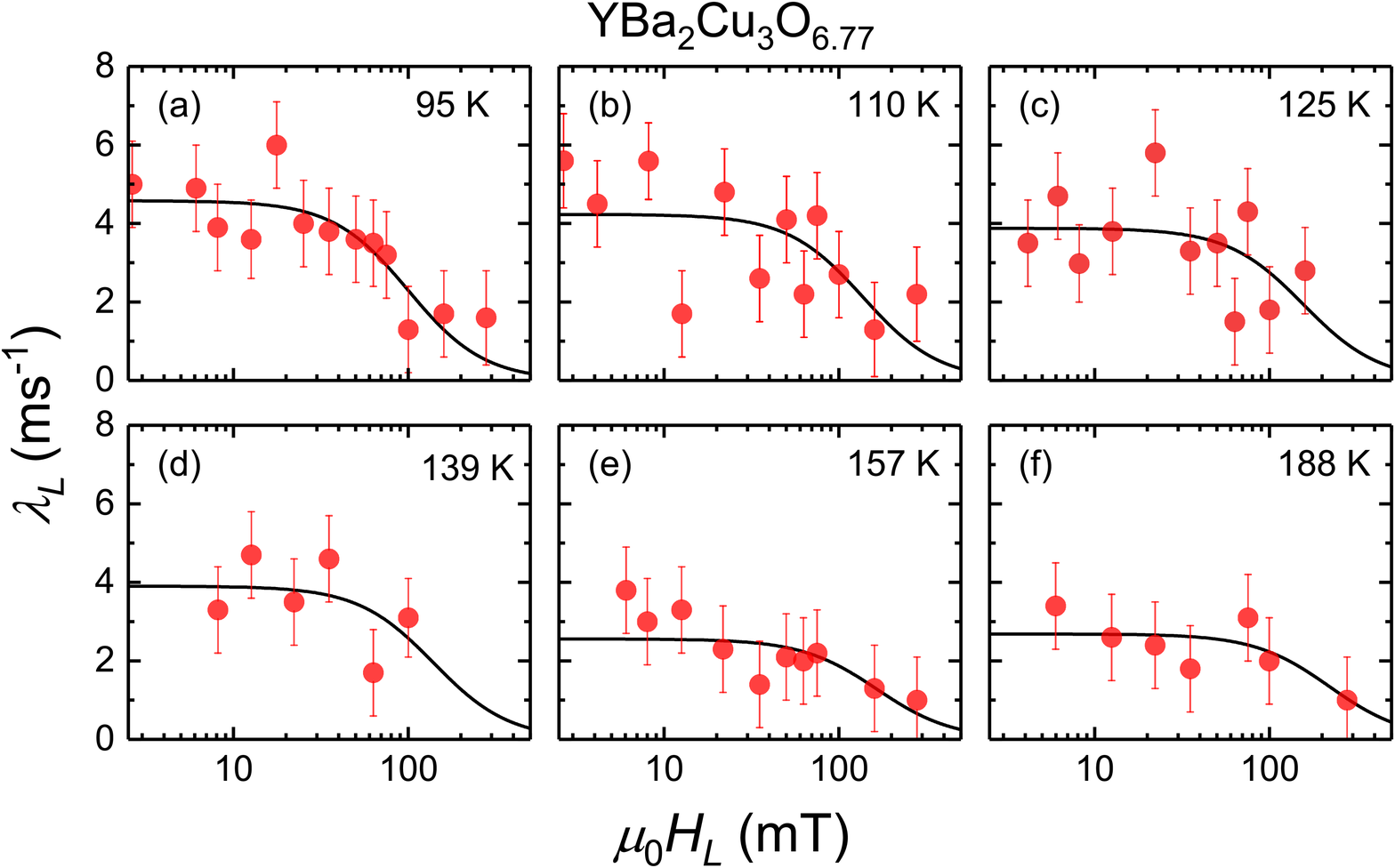}
 \caption{Dependence of dynamic $\mu^+$ relaxation rate~$\lambda_L$ on longitudinal field~$H_L$ in \YBCOS, $T > T_{c}$. Curves: fits of Eq.~(\ref{eq:Redfield}) to the data.}
 \label{fig:6p77DRR}
 \end{center}
\end{figure}

Figure~\ref{fig:6p77DRR} 
shows the field dependence of $\lambda_L$ for $y = 6.77$ at a number of temperatures. Fits to Eq.~(\ref{eq:Redfield}) are shown for $95~\text{K} \leqslant T \leqslant 188~\text{K}$. The latter is higher than $T_\mathrm{mag}$ (Table~\ref{tab:temps}), but all temperatures are within the range of $T^\ast$ values. The half-widths of the fit curves which, as noted above, are measures of $1/\tau_c$, are of the order of 100~mT, corresponding to $\tau_c \approx 10$~ns. At 157~K and above there is a significant decrease in the relaxation rate at low fields, possibly due to a spatially inhomogeneous distribution of $T^\ast$ in the sample.

Corresponding results for $y = 6.83$ are shown 
\begin{figure} [ht]
 \begin{center}
 \includegraphics[clip=,width=0.5\textwidth]{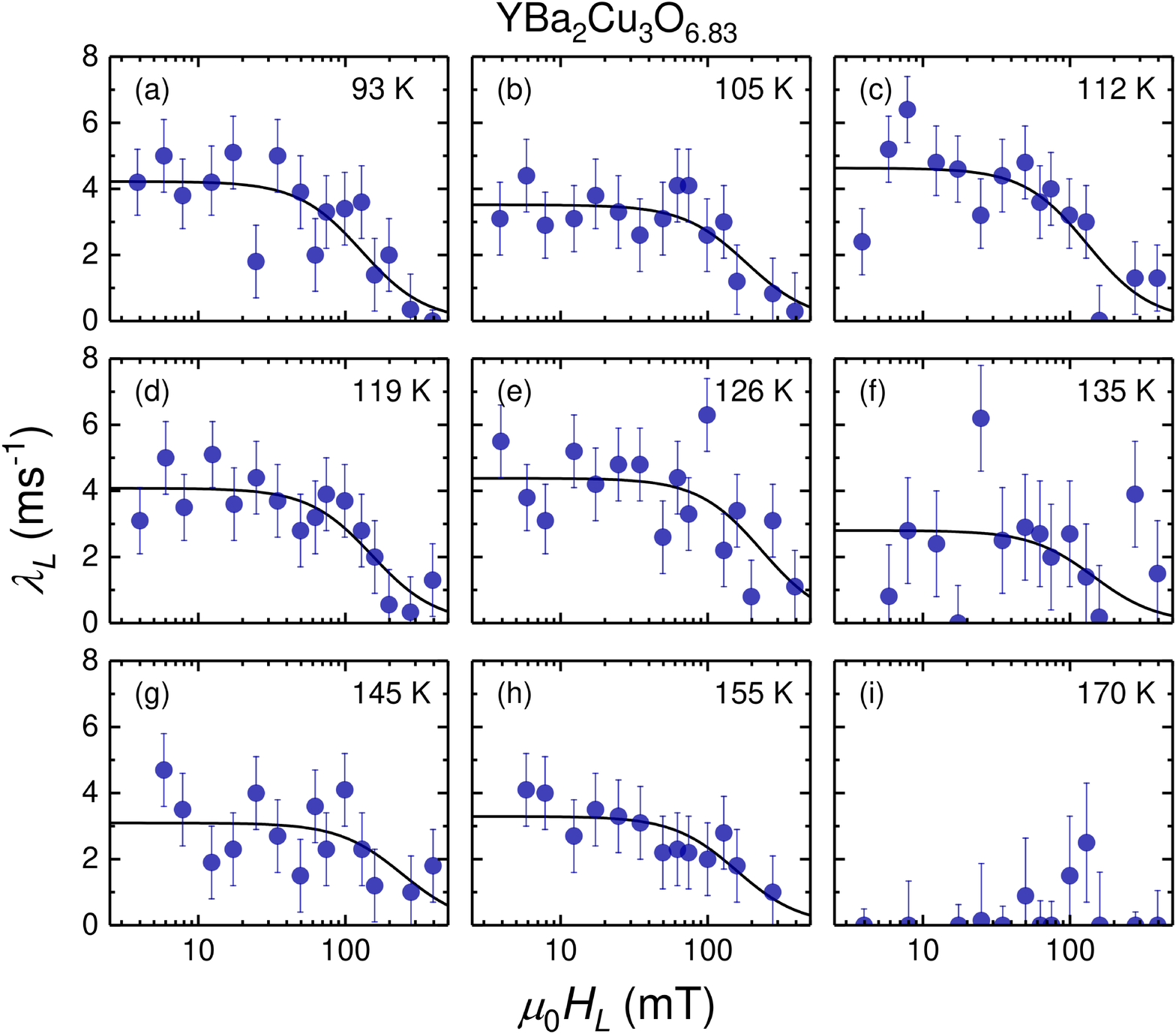}
 \caption{Dependence of $\lambda_L$ on $H_L$ in \YBCOE, $T > T_{c}$. Curves: fits of Eq.~(\ref{eq:Redfield}) to the data.}
 \label{fig:6p83DRR}
 \end{center}
\end{figure}
in Fig.~\ref{fig:6p83DRR}. Redfield field dependence of $\lambda_L(H_L)$ is again observed for temperatures up to and slightly above $T_\mathrm{mag}$, but at 170~K (above the range of $T^\ast$) the rate has fallen to zero within errors. This indicates a decrease of either $B\mathrm{_{loc}^{rms}}$ or, alternatively, of $\tau_c$ at constant $B\mathrm{_{loc}^{rms}}$, since as previously noted the area under the Redfield curve is independent of $\tau_c$. 

\begin{figure} [ht]
 \begin{center}
 \includegraphics[clip=,width=0.5\textwidth]{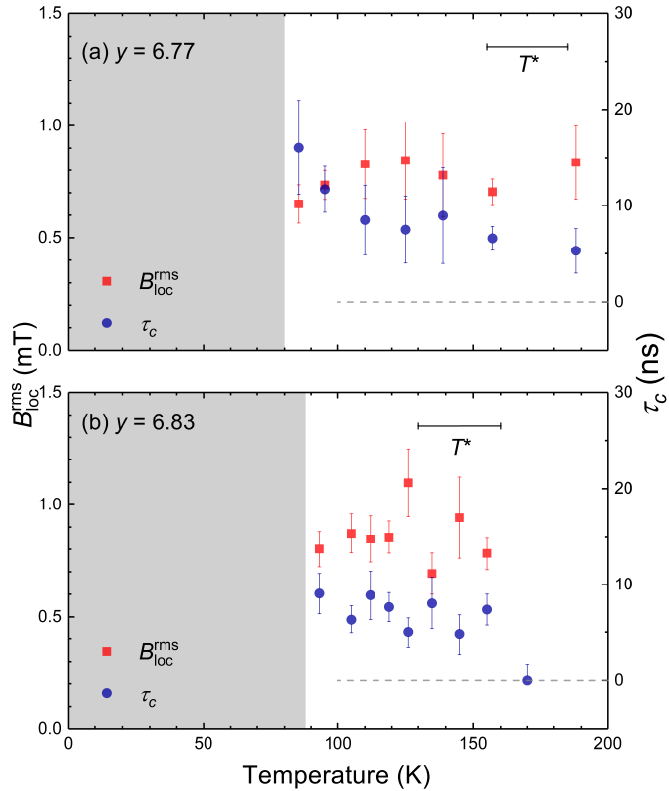}
 \caption{Temperature dependencies of $\mu^+$ rms local fields~$B\mathrm{_{loc}^{rms}}$ and correlation times~$\tau_{c}$ in \YBCO. (a): $y$ = 6.77. (b): $y$ = 6.83. Shaded areas: superconducting phase. Horizontal bars: ranges of pseudogap onset temperatures~$ T^\ast$.}
 \label{fig:OP}
 \end{center}
\end{figure}

Figure~\ref{fig:OP} 
shows the temperature dependencies of $B\mathrm{_{loc}^{rms}}$ and $\tau_{c}$ obtained from the fits. There are not enough temperature points to resolve the peak in $\lambda_L(T)$ observed previously~\cite{Zhang2018}. There is a slight and statistically marginal increase of $\tau_c$ with decreasing temperature for both dopings. For $y = 6.83$, $T = 170$~K, essentially no relaxation was observed within errors [Fig.~\ref{fig:6p83DRR}(i)]. Assuming a roughly temperature-independent $B\mathrm{_{loc}^{rms}}$, this and the consequent absence of a high-field cutoff yield $\tau_c = 0(1)$~ns. Results for \YBCOE, $T = 93$~K, differ slightly from those previously reported~\cite{Zhang2018}, most likely due to less uncertainty in the present data. We note that if fluctuating magnetism exists above $T^\ast$~\cite{Gronsleth2009}, then $B\mathrm{_{loc}^{rms}}$ is nonzero there and thus is not an order parameter in the pseudogap phase.

As has been previously noted~\cite{Zhang2018}, values of $B\mathrm{_{loc}^{rms}} \approx 1$~mT (Fig.~\ref{fig:OP}) are consistent with calculated local fields from ${\sim}0.1\mu_B$ IUC magnetic moments observed in polarized neutron diffraction in \YBCO~\cite{Fauque2006, Li2008, ManginThro2015}. The fluctuations would not affect the latter experiments: the consequent broadening~$\hbar/\tau_c$ of the diffraction pattern is ${\sim}0.1~\mu$eV, which is much smaller than the experimental energy resolution~$\sim$1~meV~\cite{CMV2014}.

As mentioned above, a series of \msr\ studies~\cite{sonier2001science, sonier2002diffusion, Sonier2009, Sonier2012, Pal2018} reported conflicting results on the nature of the detected magnetic fields in the pseudogap states of several hole-doped cuprates. Concerns were raised~\cite{Sonier2017} whether the putative dynamic field is associated with charge inhomogeneity or muon diffusion effects. However, it was shown~\cite{Zhang2018, Zhang2017} that these issues are not related to the maxima in the $\mu^+$ relaxation rates at $T_\mathrm{mag} \sim T^\ast$. We note that much of our pre\-vi\-ous\-ly-reported data~\cite{Zhang2018} and all of the present results were obtained in longitudinal fields strong enough to decouple nuclear dipolar relaxation~\cite{Hayano1979}, and are thus not subject to a recent critique~\cite{Gheidi2020}.

Information relevant to potentially corroborating NMR or NQR experiments can be extracted from our result $\lambda(H_L{=}0) = 2(\gamma_\mu B\mathrm{_{loc}^{rms}})^2\tau_c \approx 1~\text{ms}^{-1}$ (Figs.~\ref{fig:6p77DRR} and \ref{fig:6p83DRR}). To estimate nuclear spin-lattice relaxation rates~$1/T_1$, we assume fluctuating fields at nuclear sites with similar magnitudes and correlation times as reported here, and therefore scale our \msr\ results by a factor $(\gamma_\mathrm{nuc}/\gamma_\mu)^2$, where $\gamma_\mathrm{nuc}$ is the nuclear gyromagnetic ratio, to obtain estimates~$1/T_1 ^\mathrm{(est)}$ of NMR rates. These are compared in Table~\ref{tab:nuc} with experimental values~$1/T_1 ^\mathrm{(exp)}$ at $\sim$100~K for $^{63}$Cu, $^{137}$Ba $^{17}$O, and $^{89}$Y NMR in \YBCO~\cite{Walstedt2018}.
\begin{table}
\caption{\label{tab:nuc} Muon and nuclear gyromagnetic ratios~$\gamma_\mu$ and $\gamma_\mathrm{nuc}$, experimental relaxation rates~$1/T_1 ^\mathrm{(exp)}$ in \YBCO\ at $\sim$100~K (Ref.~\cite{Walstedt2018}), and estimated rates~$1/T_1 ^\mathrm{(est)}$ from pseudogap magnetic fluctuations.}
\begin{ruledtabular}
\begin{tabular}{cccccc}
 & $\mu^+$ & $^{63}$Cu & $^{137}$Ba & $^{17}$O & $^{89}$Y \\
\colrule
$\gamma_{\mu,\mathrm{nuc}}~(10^7\text{s}^{-1}~\text{T}^{-1})$ & 85.156 & 7.1088 & 2.988 & $-$3.6279 & $-$1.3155 \\ 
$1/T_1 ^\mathrm{(exp)}~(\text{s}^{-1})$ & $\sim$1000 & $\sim$2000 & $\sim$20 & $\sim$30 & $\sim$0.02 \\ 
$1/T_1 ^\mathrm{(est)}~(\text{s}^{-1})$ & -- & 7.0 & 1.2 & 1.8 & 0.2 \\
\end{tabular}
\end{ruledtabular}
\end{table}
It can be seen that, with the exception of $^{89}$Y, the estimated NMR rates are more than an order of magnitude smaller than the measured rates, making observation of the pseudogap fluctuations difficult. Furthermore, the NMR rates would suppressed by the Redfield field dependence [Eq.~(\ref{eq:Redfield})] for applied fields greater than a few tesla. In the loop-current scenario observation would also be difficult for $^{89}$Y NMR; the Y site in the \YBCO\ crystal structure is symmetric with respect to oppositely-directed IUC loop currents, and the local fields there would cancel~\cite{Mounce2013}. Thus the present results are not in conflict with the absence of NMR evidence for the slow fluctuations.

It is intriguing that similar unusual magnetic fluctuations have been observed via Redfield field dependence of $\mu^+$ dynamic relaxation in the ``hidden order'' phase of Sr$_2$Ir$_{1-x}$Rh$_{x}$O$_{4}$~\cite{Tan2019}. The hole-doped cuprates and the Rh-doped iridates share similar crystal symmetry and similarity in electronic structure and magnetic order geometry~\cite{Kim2014, Zhao2016}, and neutron diffraction experiments find evidence for TRS breaking~\cite{Jeong2017}. Both $\tau_{c}$ and $B\mathrm{_{loc}^{rms}}$ are of the same magnitude as in \YBCO, suggesting that slow spin dynamics might have the same origin in both systems.

In conclusion, we have measured the temperature dependence of the rms dynamic local fields~$B\mathrm{_{loc}^{rms}}$ and correlation times~$\tau_c$ associated with slow magnetic fluctuations in \YBCOMS, using longitudinal-field \msr\@. Although \msr\ does not yield direct information on the spatial structure of the fluctuating magnetization, the consistency of the magnitude of $B\mathrm{_{loc}^{rms}}$ with the IUC moment values from polarized neutron diffraction experiments is evidence that the fluctuating fields arise from IUC moments. The weak temperature dependencies of $B\mathrm{_{loc}^{rms}}$ and $\tau_c$ below $T_\mathrm{mag}$ are anomalous, since in conventional magnetically-ordered phases both quantities decrease with decreasing temperature below the transition. The observed behavior is reminiscent of the persistent spin dynamics (PSD) observed in spin ices and spin liquids~\cite{[{See, for example, }]Yaouanc2015, *Ding2019}. Pseudogap-phase PSD has been attributed to quantum size effects in disordered loop-current domains~\cite{CMV2014}; an alternative scenario might involve a macroscopically degenerate ground state. The long but finite correlation times are perhaps conceptually similar to long-range order in the presence of long but finite correlation lengths~\footnote{C.~M. Varma, private communication}. More work is needed to understand this situation.

\begin{acknowledgments}

We wish to thank P. Bourges, R. Kadono, Y. Matsuda. and C.~M. Varma for fruitful discussions. We are grateful to D.~J. Arseneau and B. Hitti of TRIUMF, the staff of J-PARC ARTEMIS (2017B0024), and the ISIS Cryogenics Group for valuable help during the \msr\ experiments. We also thank D.~C. Peets for suggestions on crystal preparation. This work was funded by the National Research and Development Program of China, Nos.~2017YFA0303104, 2016YFA0300503, and 2016YFA0300403, the National Natural Science Foundations of China, No.~11774061, Shanghai Municipal Science and Technology Major Project (Grant No.~2019SHZDZX01), the U. S. National Science Foundation, Nos.~DMR/PREM-1523588, HRD-1547723 and DMR-1905636, and by the Academic Senate of the University of California, Riverside.

\end{acknowledgments}

%


\end{document}